\documentclass[journal,comsoc,twocolumn]{IEEEtran}
\usepackage[T1]{fontenc}
\usepackage{cite, framed, soul}
\usepackage{amsmath,amssymb,amsfonts,amsthm}
\usepackage{algorithm2e}
\usepackage{graphicx}
\usepackage{tikz}
\usepackage{textcomp}
\usepackage{xcolor,bm,epstopdf}
\definecolor{shadecolor}{rgb}{0,1,1}
\usepackage{floatrow}
\usepackage[caption=false]{subfig}
\ifCLASSINFOpdf
\else
\fi
\usepackage[cmintegrals]{newtxmath}
\begin{document}
%
\title{\huge Smart Wireless Environment Enhanced Telecommunications: A Network Stabilisation Paradigm for Mobile Operators}
%
%
%

\author{Yangyishi~Zhang,~\IEEEmembership{Senior~Member,~IEEE},
			Khethiwe~Mhlope,
			Aaron~Walker,
			and~Fraser~Burton
        \thanks{Yangyishi Zhang, Khethiwe Mhlope, Aaron Walker, and Fraser Burton are with Research and Network Strategy, BT Group (Email: \{yangyishi.zhang, khethiwe.mhlopeziwenjere, aaron.walker, fraser.burton\}@bt.com), (Corresponding Author: Yangyishi Zhang). This work is in part supported by Horizon Europe 2020 MSCA DTN Grant 101072924, and the EPSRC Grant EP/X031977/1.}}

%



\maketitle

\begin{abstract}
Due to the uncontrolled and complex real-life radio propagation environments, Claude Shannon's information theory of communications describes fundamental limits to state-of-the-art 5G radio access network (RAN) capacity, with respect to fixed radio resource usage. Fortunately, recent research has found that a holographic metasurface-based new physical layer architecture may hold the key to overcome these fundamental limits of current mobile networks under a new paradigm, smart wireless environment (SWE), where the long-standing challenge of mobile communications, \emph{fading channel hostility}, may be solved, leading to a step-change boost in network performance and user experience.

Despite recent research activities in SWE, the best way to implement it as a network operator remains an open challenge. In this industrial review, we adopt a novel yet realistic \emph{mobile channel stabilisation} perspective for network operators to understand this paradigm shift. More specifically, we provide a technical analysis of the synergy between key next-gen mobile network enablers, e.g., holographic metasurface, wireless sensing, and machine intelligence, as well as of how this synergy leads to a robust future RAN architecture. Against the as yet unclear theoretical boundaries and low technology readiness level (TRL) of SWE enhanced telecommunications, we conclude by identifying critical challenges in future commercial deployments.
\end{abstract}


%
\IEEEpeerreviewmaketitle

\begin{figure*}[htbp]
\center
	\includegraphics[trim={0 2.5cm 0 0},clip,width=0.75\textwidth]{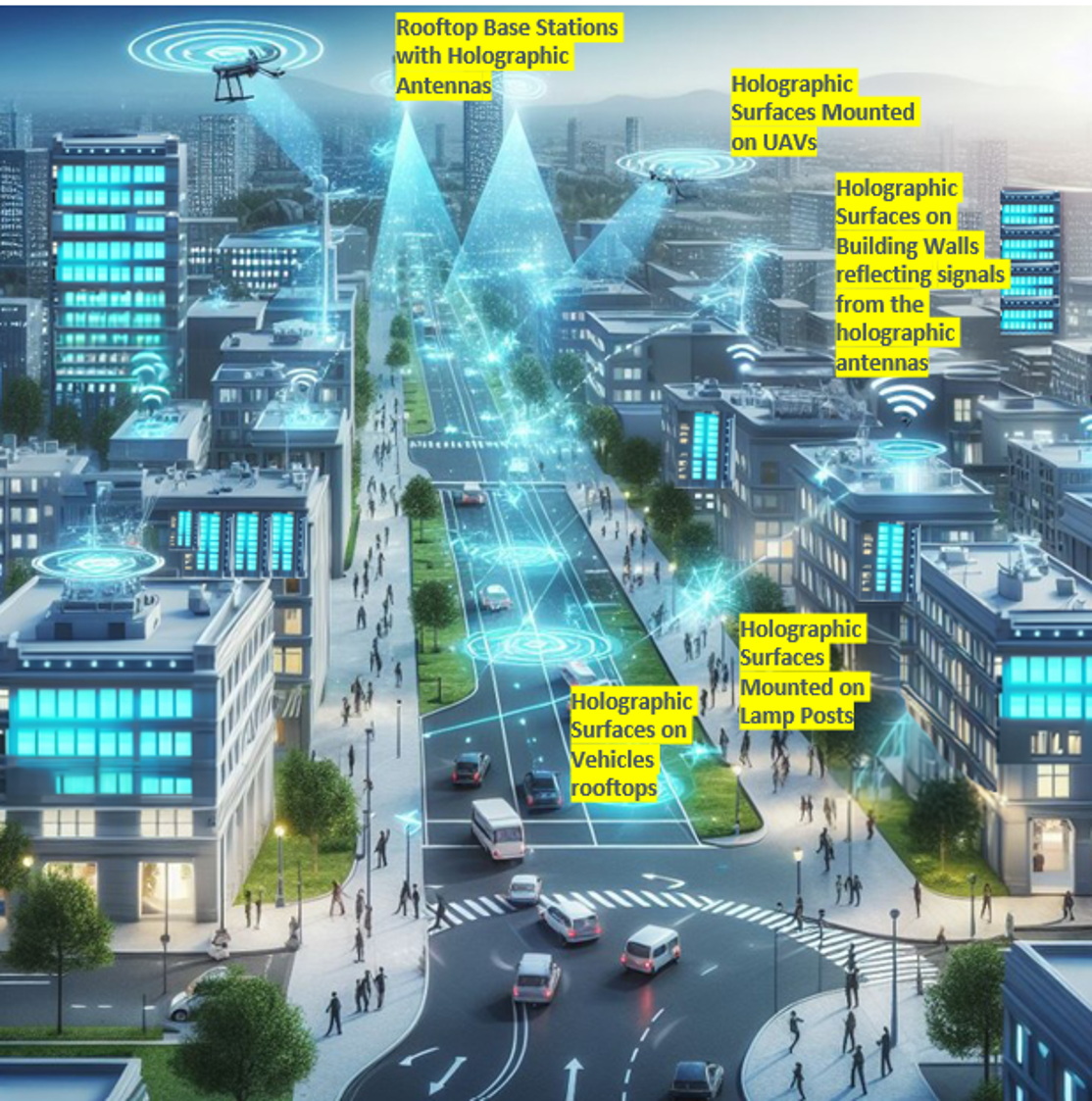}
	\caption{\label{fig:swe_city} Urban smart wireless environment (SWE) empowered by holographic metasurfaces. Holographic surfaces can fulfil the role of transceiver antenna arrays, RF reprogrammable reflectors, or simultaneously both.}
\end{figure*}
\section{Introduction}
\IEEEPARstart{D}{ue} to the escalating relevance of digital business and the Internet of Things (IoT), a faster, more reliable, and more efficient next-gen mobile network is expected to have a pivotal role in Industry 4.0 and beyond. The latest Ericsson report showed that the number of worldwide 5G mobile subscriptions will exceed 5.3 billion in 2029, and 600 million new subscriptions were gained in 2023 alone.  Use cases that demand powerful connectivity solutions, e.g., video conferencing, cloud gaming and ultra-high-definition (UHD) streaming, have taken over 73$\%$ of data traffic in the mobile market by the end of 2023. Going beyond 5G, it is critical to identify technological game changers that can drive the next mobile industry revolution to accommodate the continuous rapid growth of user demand in the future.

In the decades following Shannon's initial 1948 publication of the basic principles, information theory and our understanding of wireless communications have been greatly enriched. The dominant physical layer challenge that all mobile networks face is the uncontrolled natural environment which dictates the reflection and scattering of radio waves. The physical world is dynamic and complex so the mobile channel between transmitter and receiver distorts a signal in non-trivial ways. This introduces fundamental limits to the performance of a mobile network under fixed radio resource usage, e.g., transmit power and bandwidth. However, with the advent of holographic metasurfaces \cite{ref:Gong2024holo}, wireless sensing empowered communication \cite{ref:Liu2022integrated}, and a new era of artificial intelligence (AI), future mobile networks will be able to interact with the natural radio environments in a much more proactive way, under an emerging paradigm commonly referred to as the Smart Wireless Environment (SWE) \cite{ref:DiRenzo2020smart} (Fig. \ref{fig:swe_city}).

As the enabler of network access, mobile operators are at the forefront of realising innovative system engineering approaches to enable ubiquitous connectivity for people, devices and things. At the point of writing, SWE-related research areas, holographic metasurfaces, wireless sensing and AI, have reached unprecedented popularity in publications, funded projects, and started to gain traction in standardisation studies. Yet, the ultimate size-of-prize and real-world systematic implementation approaches for the benefits of service providers and customers are still open questions.

In this industrial overview, we discuss a narrow-sense, pragmatic approach of SWE-enhanced telecommunications (SWEET) that focuses on the delivery of network access to users. The proposed approach empowers use cases that exploit the fundamental physical properties of wireless environments to achieve network stabilisation. We will start with a holistic overview of wireless environments and channel stabilisation. Then, with the aid of an exemplary framework, we give our technical analysis of the inherent synergy between holographic metasurface, wireless sensing, and the current generation of AI, as well as how they will collaborate in stabilising the physical network. Finally, we identify both theoretical and practical open challenges that impede the commercialisation progress of SWEET frameworks.

\section{Mobile Channel and Stabilisation}
For a mobile network, the wireless environment is the physical background where radio wave propagation takes place, while a mobile channel refers to the volatile amplitude and phase distortions experienced by a radio signal when it travels from its transmitting location to the receiving location. In a real-world setting, the hostility of mobile channels is decided by the complex dynamics of background scatterers and reflectors in the wireless environment, e.g., vehicles, trees, people, etc., and also by any variations of the transceivers themselves. To transform the uncontrollable natural wireless environment into a self-optimising component of the mobile network, it is necessary to first revisit the key characteristics of natural mobile channels, and then investigate a unified channel stabilisation approach to mitigate their negative impact.

\subsection{Fundamental Channel Characteristics}
\subsubsection{Scales of Variability} \label{fcc_scale}
The behaviours of a mobile channel differ significantly based on the scale of user observations. This broadly results in three kinds of channel variations: Path loss, shadowing, and fading. Path loss, normally observable over long distances, is a deterministic process of signal attenuation that depends on both the propagation medium and background obstacles. Meanwhile, shadowing is a relatively slow-varying process observed at moderate propagation distances, usually caused by large blockage in the line-of-sight (LoS) propagation path. On a even smaller scale at a few wavelengths' distance, fading becomes the most observable type of channel variation, which appears mostly stochastic. While path loss and shadowing give crucial insight about mid to long-term power budget optimisation across RAN transmission points such as the 5G gNodeBs, fading predominantly accounts for the hostility of mobile channels.

\subsubsection{Characteristic Dimensions}
\begin{figure*}[htbp]
    \begin{centering}
    \subfloat[]{\resizebox{0.6\textwidth}{!}{%
\begin{tikzpicture}[scale = 1,
          level/.style={thick, blue},
          virtual/.style={thick, dashed, red},
          trans/.style={thick,->,shorten >=2pt,shorten <=2pt,>=stealth, line width = 3pt},
          classical/.style={thin,double,<->,shorten >=4pt,shorten <=4pt,>=stealth}
        ]  
    \tikzstyle{every node}=[font=\huge]
    \node[font=\huge] at (45.75,19.25) {$h(t, f_c, \Omega)$};
    \draw[trans] (47.5,12.5) -- (55,12.5);
    \draw[trans] (55,12) -- (47.5,12);
    \node[font=\huge] at (45.5,12.25) {$H(f_D, f_c, \Omega)$};
    \draw[trans] (47.5,19.5) -- (55,19.5);
    \draw[trans] (55,19) -- (47.5,19);
    \node [font=\huge] at (56.75,19.25) {$g(t, \tau, \Omega)$};
    \node [font=\huge] at (57,12.25) {$G(f_D, \tau, \Omega)$};
    \node [font=\huge] at (49,17) {$\phi(t, f_c, x)$};
    \node [font=\huge] at (53.5,17) {$\delta(t, \tau, x)$};
    \node [font=\huge] at (49,14.35) {$\Phi(f_D, f_c, x)$};
    \node [font=\huge] at (53.6,14.35) {$\Delta(f_D, \tau, x)$};
    \draw[trans] (46.5,18.25) -- (47.5,17.25);
    \draw[trans] (47.75,17.5) -- (46.75,18.5);
    \draw[trans] (46.5,13) -- (47.5,14);
    \draw[trans] (47.75,13.75) -- (46.75,12.75);
    \draw[trans] (54.75,13.75) -- (55.75,12.75);
    \draw[trans] (56,13) -- (55,14);
    \draw[trans] (54.75,17.5) -- (55.75,18.5);
    \draw[trans] (56,18.25) -- (55,17.25);
    \draw[trans] (50.5,17.25) -- (52,17.25);
    \draw[trans] (52,16.75) -- (50.5,16.75);
    \draw[trans] (50.75,14.5) -- (52,14.5);
    \draw[trans] (52,14) -- (50.75,14);
    \draw[trans] (48.5,16.25) -- (48.5,15);
    \draw[trans] (49,15) -- (49,16.25);
    \draw[trans] (53,16.25) -- (53,15);
    \draw[trans] (53.5,15) -- (53.5,16.25);
    \draw[trans] (45,18.5) -- (45,13);
    \draw[trans] (45.5,13) -- (45.5,18.5);
    \draw[trans] (57,18.5) -- (57,13);
    \draw[trans] (57.5,13) -- (57.5,18.5);
    \node [font=\huge] at (59.25,18.75) {$t$};
    \node [font=\huge] at (59.25,17.75) {$f_D$};
    \node [font=\huge] at (59.25,15.25) {$\tau$};
    \node [font=\huge] at (59.25,16.5) {$f_c$};
    \node [font=\huge] at (59.25,13.25) {$\Omega$};
    \node [font=\huge] at (59.25,14.25) {$x$};
    \node [font=\LARGE] at (61,18.75) {Time};
    \node [font=\LARGE] at (61.25,17.75) {Doppler};
    \node [font=\LARGE] at (61.5,16.5) {Frequency};
    \node [font=\LARGE] at (61,15.25) {Delay};
    \node [font=\LARGE] at (61,14.25) {Space};
    \node [font=\LARGE] at (61,13.25) {Angle};
\end{tikzpicture}\label{fig:cphys}}}
    
    \subfloat[]{\includegraphics[width=0.33\textwidth]{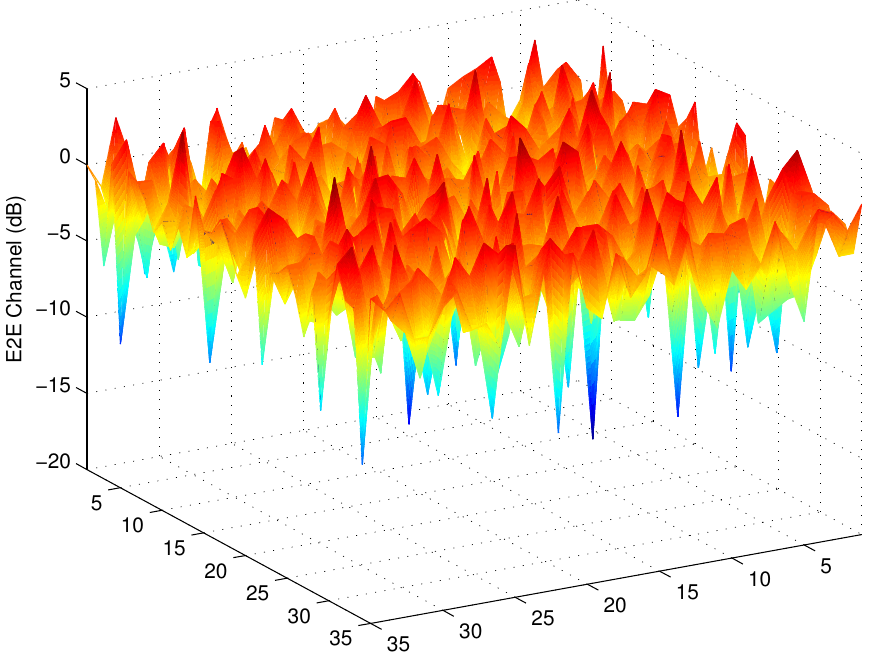}\label{fig:rawc}}
	\subfloat[]{\includegraphics[width=0.33\textwidth]{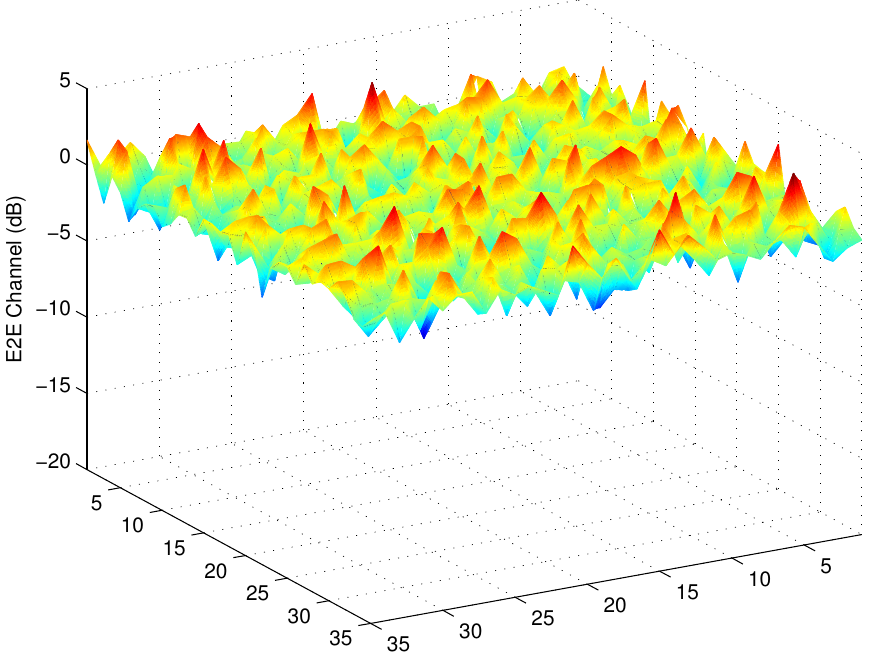}\label{fig:gsc}}
	\subfloat[]{\includegraphics[width=0.33\textwidth]{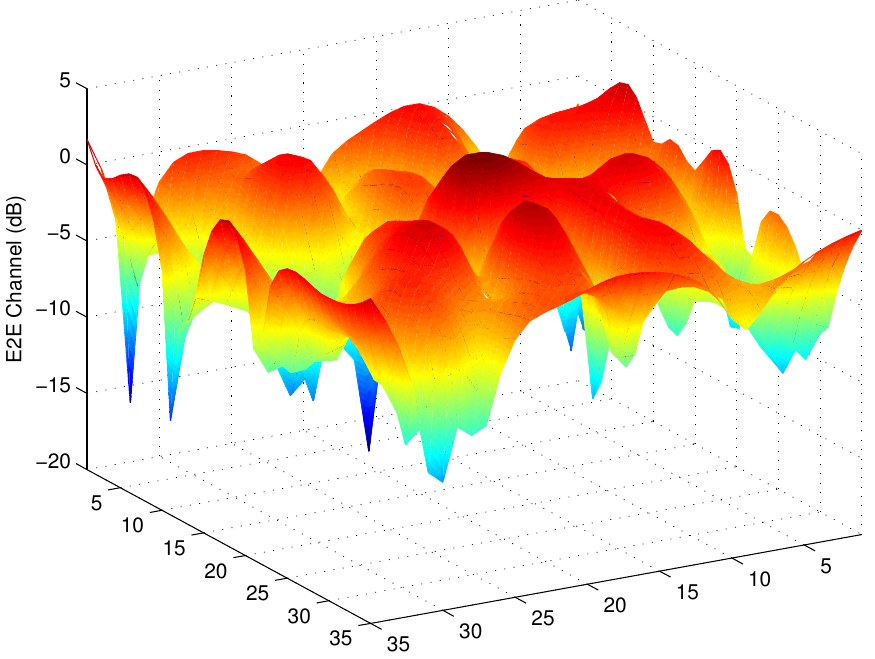}\label{fig:lsc}}
    \caption{\label{fig:stability} E2E channel representation in characteristic dimensions: (a) Fourier transform relationship between characteristic (time, frequency, space) and physical (Doppler, delay, angle) dimensions, (b) Unstable fading channel response, (c) Globally stable fading channel with reduced dynamic range, and (d) Locally stable fading channel with increased correlation.}
    \end{centering}
\end{figure*}
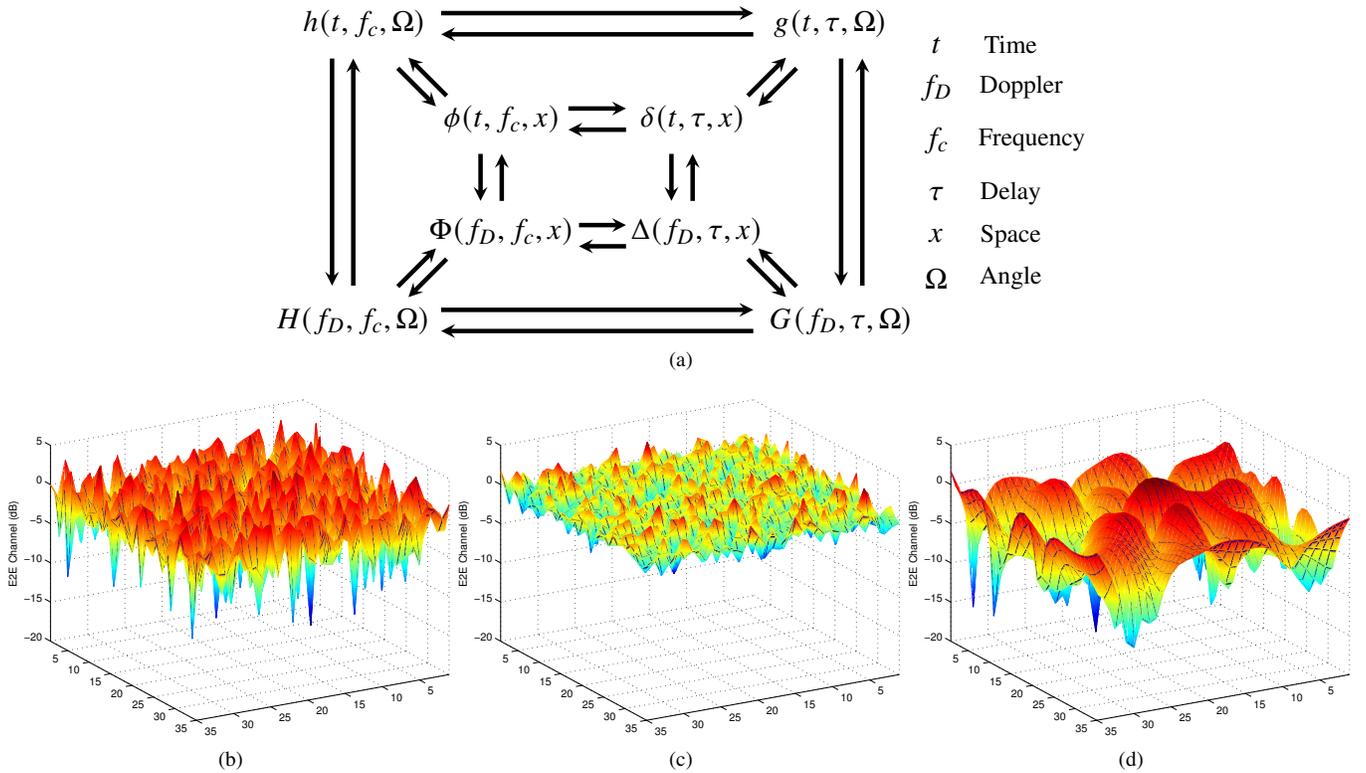

Physical characterisation of fading channels reveals important multi-dimensional features that can be exploited in network engineering. It is currently well-known a fading mobile channel has at least three characteristic dimensions \cite{ref:Cheng2022channel}: Time, frequency, and space, which are associated by the Fourier Transform to their respective physical parameters (Fig. \ref{fig:cphys}), namely (a) Doppler frequency, caused by relative motion, (b) multipath delay, caused by asynchronous reception of scattered signals, and (c) angular spread, caused by spatially separated arrivals/departures of signal copies. Typically, the stability of a fading channel is portrayed by \emph{coherence intervals}, which refer to blocks of quasi-static or highly correlated channel response in one or more of the characteristic dimensions. We note that some systems introduce new characteristic dimensions to fading channels for multiple access purposes, e.g., spreading codes. While the additional dimensions are indeed visible to transceivers, they are not induced by the environment. Hence we will not focus on them for the rest of our discussions.

\subsection{Fading Channel Stabilisation}
Shannon's equation formulates the relationship between a communication system's performance limit, its radio resource usage, and its operating environment: The better the \emph{environment} is, the more traffic a mobile network can accommodate under fixed radio resource usage. In pursue of this holy grail, mobile networks have historically adopted a variety of information processing approaches to improve the perceived end-to-end (E2E) channel condition. Classically, these approaches have only been performed at the transmitter-side and/or receiver-side, which can be broadly categorised as three categories: Error correction, beamforming, and multicarrier. While error correction is mainly information-based without explicit dependence on the physical environment, both beamforming and multicarrier exploit the physical characteristics of wireless environments to improve the perceived E2E channel. Specifically, multicarrier modulation segments one or more of the channel's characteristic dimensions to operate under many quasi-static E2E subchannels, while beamforming relies on estimating the channel state information (CSI) in characteristic dimensions to remove any learned variations.

In this review, we consider fading channel stabilisation to be a natural evolution and consolidation of the state-of-the-art measures, built on the foundations of multicarrier and beamforming. It refers to a regime where \emph{the perceived E2E channel condition is flattened in one or more of the characteristic dimensions}. This concept is portrayed in Fig. \ref{fig:stability}, which characterises both global and local channel stabilities. Different from the scale over propagation distance discussed in Sec. \ref{fcc_scale}, here we define the global and local scales of a \emph{fading} channel within its characteristic dimensions as follows.

\subsubsection{Global Channel Stabilisation}
Multiple-input-multiple-output (MIMO) technologies have received remarkable real-world success in network engineering. The physical nature of MIMO is that, when sufficiently separated in space, several antenna elements each spawns an weakly correlated spatial subchannel due to different collections of scattered propagation paths. Under this assumption, MIMO may exploit the spatial \emph{diversity} shown by its subchannels to harness a \emph{high-rank} equivalent E2E channel.

The improvement of channel rank has a self-stabilising effect: In the limit of many antenna elements, an equivalent E2E MIMO channel becomes static. This well-known phenomenon, i.e., channel hardening, shows the stabilisation of fading channel over a global time scale for the global range of antenna elements. Here the idea of collecting independent subchannels to form a stable E2E global channel on the global scale is termed as global channel stabilisation.
\begin{figure*}[htbp]
\center
	\includegraphics[trim={0.5cm 3.5cm 4.8cm 1.2cm},clip, width=0.75\textwidth]{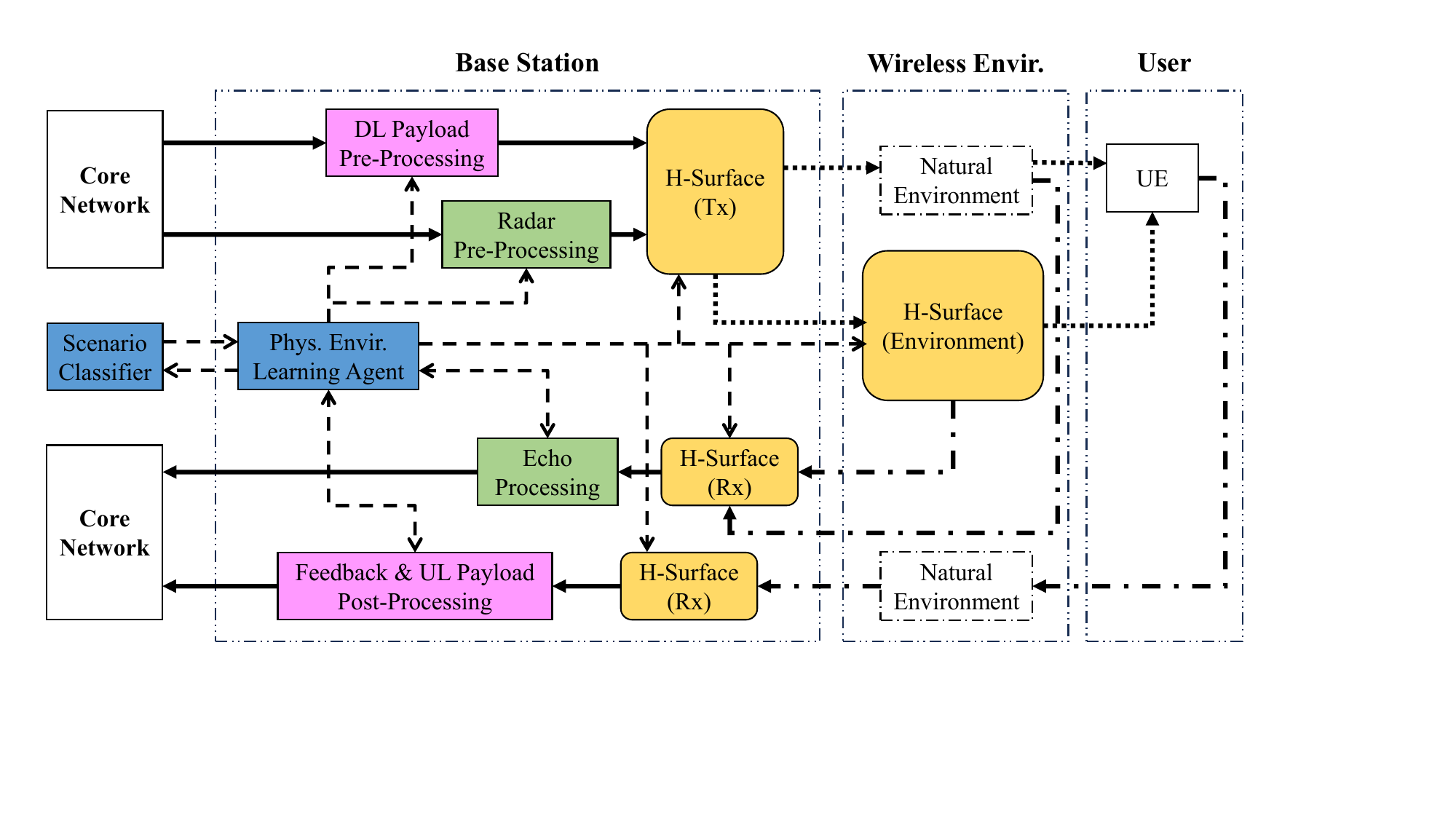}
	\caption{\label{fig:sweet} The smart wireless environment enhanced telecommunication (SWEET) framework. Adapting the quality of physical mobile channels to network stabilisation needs requires three key components: Holographic metasurface (yellow blocks) aided \emph{environment augmentation}, radar sensing (green blocks) aided \emph{environment awareness}, and machine intelligence (blue blocks) aided \emph{environment control algorithm}. SWEET fundamentally relies on centrally managed large-scale AI to classify network scenarios, exploit similarities in physical environment attributes (e.g., whether a scenario consists primarily of humans or vehicles), and configure distributed physical environment learning agents. The learning agents are low-profile AI models which directly control physical layer operations at the edge and provide necessary updates to the scenario classifier.}
\end{figure*}

\subsubsection{Local Channel Stabilisation}
The success of classical MIMO largely depends on having near-perfect CSI knowledge \emph{of individual subchannels} at the transmitter and/or receiver side, but the acquisition of CSI in existing mobile networks relies on high-complexity and resource-hungry yet inaccurate channel estimation procedures. In essence, the CSI estimated at one point in a characteristic dimension does not normally match the CSI at another point. Hence mobile networks are vulnerable, even sensitive, to channel estimation errors.

Local channel stabilisation focuses on individual local subchannels over a short local interval. Based on observable physical correlations, data analysis tools are utilised to first find or synthesise a channel response function over its characteristic dimensions, and then invert the \emph{interpolated or predicted} channel response at points adjacent to the known CSI points. Clearly, channel prediction or interpolation approaches are expected to perform the best over short intervals in the characteristic dimensions where strong correlation can be captured. Statistically uncorrelated channels are difficult to predict or interpolate accurately, which may instead degrade network performance further. However, as channel response reflects both physical wave propagation and the dynamics of macroscopic objects, statistically uncorrelated channels do not exist in the real world, meanwhile accurate insights about the wireless environment's physical realities are critical for achieving local stabilisation gain.

\section{The SWEET Framework} \label{SWEET}
The SWE paradigm adopts radically new network engineering approaches. By directly intervening in background wave propagation with the aid of environment augmentation, future mobile networks will gain the power to tailor physical channel response \cite{basar_reconfigurable_2020}, stabilise system operations, and improve user quality of experience. Hence, for the purpose of mobile service delivery, fading channel stabilisation becomes a core use case of SWEET in terms of improving network reliability, reducing manual cost, and simplifying operations for existing networks. Behind key components of the envisioned framework (Fig. \ref{fig:sweet}), we will discuss their respective R$\&$D landscape and an operator's deployment considerations.

\subsection{Environment Augmentation: Holographic Metasurfaces}
\begin{figure*}[htbp]
    \begin{centering}
    \subfloat[]{\includegraphics[trim={1.5cm 1.5cm 2.5cm 1.5cm},clip,width=0.49\textwidth]{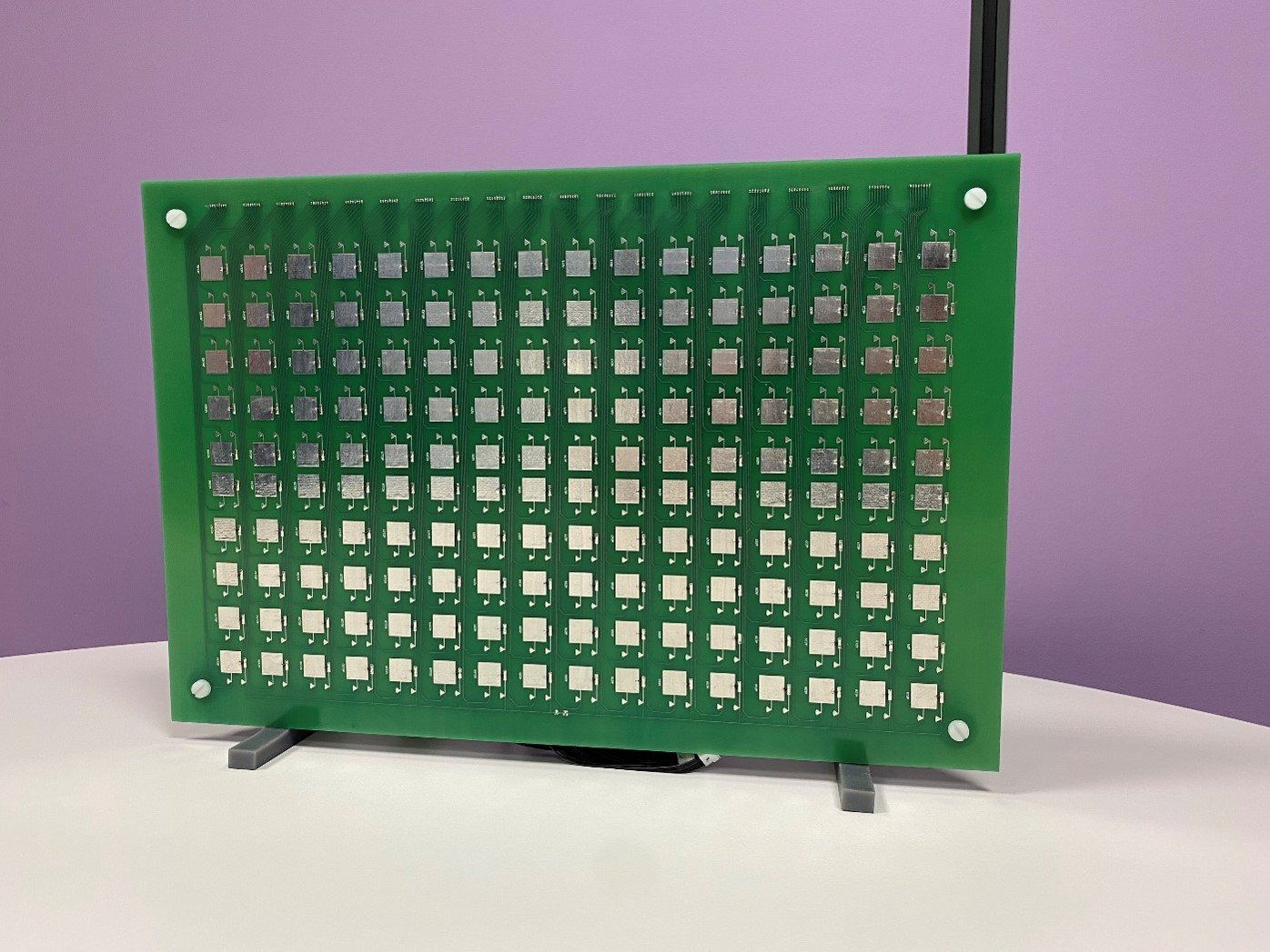}\label{fig:holof}}
	\subfloat[]{\includegraphics[trim={1.5cm 1.5cm 2.5cm 1.5cm},clip,width=0.49\textwidth]{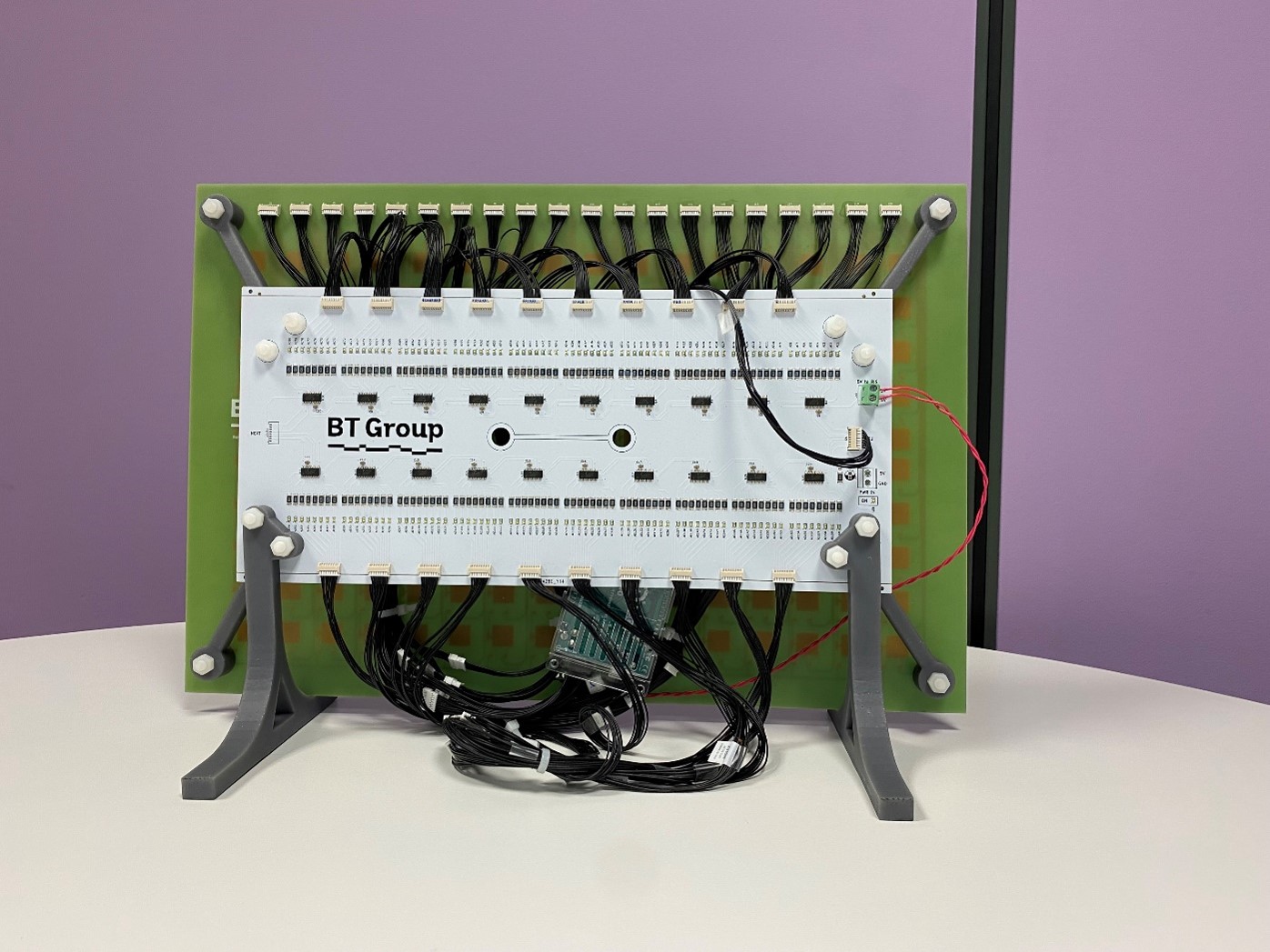}\label{fig:holob}}
    \caption{\label{fig:holos} Experimental prototype of a reflective holographic metasurface (IRS/RIS). (a) Unit elements and surface tiling \cite{trichopoulos_design_2022}, (b) Control logic.}
    \end{centering}
\end{figure*}

In the broad context of SWE, holographic metasurface \cite{ref:Gong2024holo} refers to a family of mostly planar, electrically thin radio-frequency (RF) structures that emit \emph{programmable} RF output given a specific RF input (e.g., Fig. \ref{fig:holos}). Depending on the type of input RF source, holographic surface can fulfil a variety of roles in SWE, as either transmit array upgrades or augmentation of the wireless background, leading to somewhat chaotic naming conventions in the literature. According to their core functionalities, RF sources, and operational overheads, these surfaces are categorised in Tab. \ref{tab:holos}. Note that for environment augmentation, the holographic surfaces incur extra control overhead on top of transceiver array configuration cost, whilst solely transmissive types do not. For more advanced roles or improved performances, they can also lead to sub variants such as intelligent omni surface (IOS), i.e., a reflective and refractive holographic metasurface, and stacked intelligent metasurface (SIM), i.e., a multi-layered transceiver array.

To facilitate channel stabilisation, the SWEET framework needs an elegant overall solution for the RF frontend: both transceiver-side beamforming and environment-side augmentation should use the appropriate type of holographic surfaces. The specific choice is vendor discretionary, but versatility of installation, low power profile, low control overhead, and ease of maintenance are critical factors behind their feasibility for commercial deployment. Additionally, large aperture size is also preferable for expanding the coverage of controlled environment. However, due to the limitations of existing RF circuitry designs behind the array elements, the RF response of holographic surfaces has fundamental phase-amplitude self coupling, resulting in constrained programmability.

\subsection{Environment Awareness: Sensing-Aided Communications}
To support the superior stability of an ultra reliable SWEET framework, a key requirement is to accurately perceive physical attributes of the natural network environment, such that the physical causes of channel variations, e.g., vehicle movement, can be correctly identified, tracked, and compensated for. Environment awareness needs to be gained from sufficient topology-specific network information. Such a procedure natively benefits from wireless sensing technology. Future RAN is expected to leverage sensing under a hybrid signal processing framework, namely integrated sensing and communication (ISAC) \cite{ref:Liu2022integrated}, with the expectation to save radio resources via compact waveform designs. However, although ISAC attempts to merge sensing with communication, it still tends to incur spectral overhead for mobile operators, as radar sensing normally requires much higher frequencies beyond most 5G carrier bands. Moreover, the trade-off between sensing and communication performances is known to have fundamental limits, which will critically influence the stabilisation gain of SWEET networks.

While environment-aware communication frameworks already exist in the research literature and even in limited commercial scenarios, e.g., Wi-Fi sensing, historical designs mainly exploit distributed CSI sounding to build an information-level abstraction of the network topology, e.g., channel knowledge map or channel chart. By contrast, SWEET relies on sensing-aided communication to gain physical-level environment knowledge, such as identifying the dominant Doppler shift contributors. This choice exploits the fact that ISAC \emph{echo} and CSI \emph{feedback} carry correlated yet non-identical information about the wireless environment, due to the time gap between their receptions. Moreover, sensing is effective against environment obstacles that are unresponsive to communication signals, i.e., device-free sensing \cite{ref:Wang2018device}, which can capture more physical details in environment measurement. Finally, as remotely installed holographic surfaces, e.g., RIS, can perform localised, shared-aperture ISAC \cite{ref:Joy2024from}, a SWEET network may utilise distributed sensing to further improve its environment awareness.

\subsection{Environment Control: Large Intelligent Physical Models}
With the escalating amount of mobile users and traffic, SWEET will operate in challenging network environments. Since the physical network gains environment awareness from sensing data, a suitable stabilisation protocol is needed to extract and interpret the physical environment's significant features, usually based on long-term global dependencies and short-term local correlations in characteristic dimensions. Such a data processing methodology has natural synergy with recent breakthroughs in large-scale machine learning (ML): Developments of graphical processing unit (GPU) have empowered the commercial deployments of multi-hundred-billion parameter deep neural networks, e.g., the family of Transformer-based large-scale ML models, whose success has caused a fundamental paradigm shift among real-world businesses.
\begin{table*}[htbp]
\begin{center}
\caption{{\label{tab:holos}}Summary of Holographic Metasurfaces for SWE}
\begin{tabular}{|p{6cm}|p{4.2cm}|p{1.7cm}|p{1.5cm}|p{1.5cm}|}
\hline
 Name  & Functionality & RF Source & Energy\newline overhead & Control\newline overhead  \\ \hline
\hline
Reconfigurable intelligent surface (RIS)/\newline Intelligent reflecting surface (IRS) & Environment augmentation & Wireless & Low & Mid\\ \hline
Active (\emph{Amplify and reflect}) RIS & Environment augmentation  & Wireless & Mid & Mid  \\ \hline
Reconfigurable holographic surface (RHS)  & Transceiver array variation & Wired & N/A & Low \\ \hline
Simultaneous transmitting and reflecting (STAR) RIS  & Distributed array variation $\&$\newline Environment augmentation & Wired $\&$\newline Wireless & High & High \\ \hline
\end{tabular}
\end{center}
\end{table*}

AI-based optimisation of the mobile RF frontend predates long before the birth of Transformers. However, with the versatility of Transformer-inspired ML, a unified large physical model is envisioned to fulfil multiple key roles for channel quality improvement within SWEET. Most notably, Transformer has already been shown to radically improve the performance of channel prediction \cite{ref:Jiang2022accurate} in classical networks. Secondly, with recent advances in AI-aided video synthesis, Transformer-based diffusion models have also shown high potential in understanding basic correlations of the real world \cite{ref:Liu2024sora}. Finally, due to consistency in performance requirements and environment attributes, network scenarios can be categorised, e.g., stadiums, urban streets, etc., as individual base models for better use case-specific model training and fine-tuning.

Network intelligence is expected to be a fundamental enabler beyond 5G. However, as a physical layer control framework, it raises several practical concerns. First, the mobile physical layer has stringent latency requirements, especially for time-critical tasks such as steering autonomous vehicles, which are hard to meet by large neural networks. Moreover, training and tuning large-scale ML models knowingly incur significant energy cost, hence the ultimate energy trade-off from network stabilisation needs further experimental validations. Lastly, the elegant operations of neural networks, including the most powerful generative AI models, have highly exploitable vulnerabilities \cite{ref:Yang2024a}, e.g., prompt attack, which requires non-trivial mitigation mechanism to maintain system integrity.

\section{TRL Development Challenges}
The effectiveness of SWEET channel stabilisation has a clear vision: Holographic surfaces can remove hostile channel variations resulting from known or predictable Doppler and multipath effects \cite{basar_reconfigurable_2020}, while sensing-aided physical layer network intelligence will be responsible for accurately capturing and predicting them in real environments \cite{ref:Niu2022rethink}. However, to understand and realise its potential benefits to real-world network service delivery and user experiences, there are still challenges ahead. At the time of writing, the SWEET enablers discussed earlier have yet to advance beyond initial technology readiness levels (TRLs), as existing research investigations largely stay on simulations, i.e., TRL2, or characteristic proof-of-concepts, i.e., TRL3. The following sections will discuss critical issues related to the tangible, real-world realisation of a SWEET ecosystem for service providers, suppliers, and customers.

\subsection{Physically Consistent System Limits} 
Due to modifications of the physical infrastructure, analysing the SWEET framework requires novel modelling approaches. This is largely due to its ability to influence mobile networks in the RF domain, while classical communication models are mostly built in the baseband digital domain and they tend to incorporate unphysical assumptions. Statistical models such as the Rayleigh model are primarily for the understanding of information theory at large without having to obey laws of physics. On the other hand, existing physical models until the last few years have been largely treated in simple forms, such as RF circuits or clustered ray tracers. To truly bridge the gap between Shannon's and Maxwell's first principles, researchers recently started attempting to formulate a robust electromagnetic information theory (EMIT) framework \cite{ref:Zhu2024EMIT} for understanding the physically consistent limits of SWEET. However, EMIT is still at an early stage that awaits further development and refinement.

\subsection{System-Level Prototyping}
System-level prototyping has paramount importance in commercial development of the SWEET framework, as it provides real-world implementation insights which can also improve the quality of both theoretical models and simulated trials. For example, under realistic conditions, a SWEET system exhibits three types of fundamental correlations that tend to be unaccounted for in most known simulators and testbeds: The phase-amplitude correlation of a metasurface unit element, the mutual correlation between unit elements, and the spatial correlation between \emph{cascaded channel segments} (e.g., a trio of transmitter-RIS, RIS-receiver, and E2E direct channels). More generally, the clustered ray tracing approach many existing mobile system simulators adopt may not adequately address the SWEET framework's unique physical traits, such as the coupling of physical fading channel parameters perceived by holographic surfaces in a shared wireless environment. Hence, new developments in public datasets, system simulators, and testbed platforms are in particular need for understanding the gap between real-world implementations and physical models.

\subsection{Regulation Compliance}
While standardisation for holographic surfaces, ISAC protocols, and network intelligence is still work in progress, numerous immediate regulatory issues have already been identified \cite{ref:ofcom2024}. Most notably, as holographic surfaces will extend outside of operator-owned space, e.g., radio towers, into public environments, their designated ability to modify radio propagation may be problematic. Under the presence of users served by different network operators, the design limitations and/or imperfections of holographic surfaces may lead to either unintended modifications of competitor's services, spectral leakage into \emph{unauthorised} bands, or simultaneously both. Additionally, while sensing-aided communication is a critical design element for environment awareness, location-sensitive information gathering from public spaces might constitute a source of privacy violation unless appropriately anonymised. Moreover, due to the limited explainability of current large AI models, its central spotlight role in physical network control may cause unforeseen risks to users and operators.

\subsection{Coexistence with Current Networks}
While SWEET introduces disruptive approaches to physical network engineering, it needs to be compatible with, at least smoothly transitioned from, current-gen networks and scheduled near-future deployments. To this end, the radical change to physical network infrastructure caused by holographic surfaces will have an notable impact: Although they are intended to simply augment the wireless environment to stabilise network conditions, the design choice of directing signal propagation to relatively confined areas in a typically wide open space may potentially degrade the natural spatial diversity of existing MIMO frameworks \cite{ref:Rains2024RIS}. Additionally, while improved channel correlation (e.g., Fig. \ref{fig:lsc}) can mitigate channel estimation errors, \emph{burst residual errors} due to long-lasting bad channel conditions will become more likely. This tends to have a detrimental effect on forward error correction coding schemes, as they can only correct a limited number of errors within an encoded sequence. When said limit is exceeded, error correction schemes break down and further degrade network performance.

\section{Conclusions}
This paper discusses a practical route to value for mobile network operators to embrace the new paradigm of smart wireless environments among other emerging 6G enablers. To gain control of the hostile mobile channel and tame the detrimental fading effect caused by natural wireless environments, we propose a SWEET framework that unites holographic surface based environment augmentation, sensing aided environment awareness, and machine intelligence aided environment control. Their technical relevance and research landscapes are analysed with respect to the requirements behind mobile access service delivery. Finally, fundamental commercialisation challenges are reviewed.


%

%

\section*{Acknowledgment}
The authors would like to thank Li Lin, Liam Bussey, Samuel Winter, Minglei You, Tianrui Chen, Anas Al Rawi, Gan Zheng, Gabriele Gradoni, Marco Di Renzo, Lajos Hanzo for the insightful discussions during various stages of this work.

\ifCLASSOPTIONcaptionsoff
  \newpage
\fi



%
\bibliographystyle{ieeetran}
\bibliography{MyRefs}

\end{document}